\documentclass[twocolumn,aps,floats,showpacs]{revtex4}
\usepackage{amsmath}
\usepackage{bbm}
\usepackage{graphicx}

\begin{document}

\bibliographystyle{apsrev}

\title{Magnetoresistance of itinerant electrons interacting with local spins}

\author{G. G\'omez-Santos$^{1}$, S. Fratini$^{2}$, and F. Guinea$^{3}$}

\affiliation{$^{(1)}$Departamento de F\'{i}sica de la Materia
Condensada and Instituto Nicol\'as Cabrera, Universidad Aut\'onoma
de Madrid~28049~Madrid,~Spain. \\ $^{(2)}$Laboratoire d'Etudes des
Propri\'et\'es Electroniques des Solides, CNRS-BP166. 25, Avenue
des Martyrs, F-38042 Grenoble Cedex 9. \\ $^{(3)}$Instituto de
Ciencia de Materiales de Madrid
(CSIC),~Cantoblanco,~28049~Madrid,~Spain. }

\begin{abstract}
 Transport properties of itinerant
electrons interacting with local spins are analyzed as function of
the bandwidth, $W$, the exchange interaction, $J$, and the band
filling, $n$, near the band edge. Numerical results have been
obtained within the dynamical mean-field approximation and
interpreted with the help of analytical treatments. If spatial
correlations of the magnetic fluctuations can be neglected, and
defining $\rho ( m )$ as the magnetization dependent resistivity,
we find that $\delta \rho / \rho ( 0 ) \sim n^{-4/3}$ for $J / W ,
n \ll 1$ (weak coupling, low density limit), $\delta \rho / \rho (
0 ) \sim constant$, for $J / W \gg 1 , n \ll 1$ (double exchange,
low density limit), where $\delta \rho = \partial \rho / \partial
m^2 |_{m^2 \rightarrow 0}$. Possible limitations from ignoring
both localization effects and  critical fluctuations are also
considered.
\end{abstract}

\maketitle

\section{Introduction.}
A large variety of ferromagnetic materials can be described in
terms of a band of itinerant electrons interacting with localized
spins. Among others, we can include the manganese perovskites,
La$_{2-x}$Re$_x$MnO$_3$\cite{CVM99}, the doped pyrochlores,
derived from Tl$_2$Mn$_2$O$_7$\cite{SKM96,Metal99,Vetal02}, and
many doped magnetic semiconductors\cite{O98,P98}. Many of these
materials show unusually large values of the magnetoresistance
near the Curie temperature. The simplest description of the
electronic and magnetic properties of these compounds includes an
itinerant band of non interacting electrons, parameterized by the
bandwidth $W$, and a finite concentration of classical localized
spins, which interact with the electrons through a local exchange
term, $J$. In the following, we assume that the concentration of
spins is of the same order as the number of unit cells in the
lattice, and describe the model in terms of two dimensionless
parameters, the ratio $J/W$ and the number of electrons per unit
cell, $n$. The values of these parameters for the  pyrochlore
compounds typically satisfy $J/W , n \ll 1$, while the double
exchange model appropriate for the manganites is such that $J/W
\gg 1$, and $n$ can be of order unity. The simple description used
here is probably inadequate for the dilute magnetic
semiconductors, where the concentration of local moments is small,
and, moreover, disorder effects can lead to localization of the
electronic states\cite{CGB02}.

It has proven useful to identify general trends in the transport
properties of these materials. The resistance is typically high in
the paramagnetic phase, and decreases as the magnetization
increases. A simple parameter used to characterize the
magnetoresistance is the dimensionless ratio ${\cal C} = [ \partial \rho
( m^2 ) / \partial m^2 ] / \rho |_{m^2 \rightarrow 0}$, where
$\rho ( m^2 )$ is the magnetization dependent resistance. The
resistivity is dominated by scattering processes with wavevector
$| {\bf \vec{q}} | \sim 2 k_F$\cite{fl68}, where $k_F$ is the Fermi momentum.
The scattering arises from magnetic fluctuations, which are
described by the magnetic susceptibility, $\chi ( {\bf \vec{q}}
)$.  Assuming that $\chi ( {\bf \vec{q}} )$ can be approximated
taking into account only the low momentum critical fluctuations
which exist near the Curie temperature, one finds that $C \propto
n^{-2/3}$\cite{ML98,ML98b}.

A different regime exists at sufficiently large values of the
electronic density, when $\chi ( | {\bf \vec{q}} | = 2 k_F )$
cannot be described in terms of critical fluctuations. In a system
with a large coordination number, the properties of $\chi ( {\bf
\vec{q}} )$ for sufficiently large values of ${\bf \vec{q}}$ can
be studied using an effective medium approach, which neglects the
spatial correlations of the magnetic fluctuations. As it is well
known, this description becomes exact when the dimensionality, or
the coordination number, becomes large\cite{Getal96}. The
conductivity of models of itinerant electrons interacting with
classical spins has already been analyzed in this
limit\cite{F94,F95}. In the following, we will study numerically
the dependence of the parameter ${\cal C} $ defined above on the
density of itinerant electrons,  mostly in the regime when
$\tilde{\epsilon}_F \ll W$, where $\tilde{\epsilon}_F$ is the
Fermi energy measured from the bottom of the band, assuming
uncorrelated magnetic disorder.  We will perform numerical
calculations in the Coherent-Potential Approximation and use
simplified analytical methods to understand the numerical results.
We will find that, for $J/W \ll 1$, the band splitting due to the
magnetization leads to the divergent
behavior, ${\cal C} \propto n^{-4/3}$. On the other hand, the
magnetoresistance coefficient $ {\cal C}$ remains finite close to
the band edge in the double exchange limit $J/W \gg 1$, a fact
easily understood in terms of a magnetization-dependent bandwidth.
The paper is organized as follows. Section \ref{perturb} presents
the model and the perturbation limit for ${\cal C}$. The numerical
method is described in section \ref{cpa}. Section \ref{weak}
contains results for the weak coupling limit, with a simplified
analytical treatment and a discussion of possible limitations due
to localization effects. Results for the strong coupling limit and
a simplified explanation are found in section \ref{strong}. The
last section summarizes the main conclusions of our work and makes
contact with the experimental situation, placing out results in
the context of available compilations of magnetoresistance
data\cite{ML98,Vetal02,V02}.

\section{Model Hamiltonian. Magnetoresistance in the perturbative limit}
\label{perturb}
We consider free electrons coupled to core spins and modeled by the following
Hamiltonian:
\begin{equation}
H=\sum_{\bf{k},\sigma} \epsilon_{\bf{k}}  \: c^{\dagger}_{\bf{k},\sigma} c_{\bf{k},\sigma} -
J \sum_{i} {\bf S}_i \cdot \mbox{\boldmath$\sigma$}_i
\label{hamil}
\end{equation}
where $\sigma = \pm1$ is a spin index, $\epsilon_{\bf{k}}$ is the
band structure of carriers in the unperturbed system,
$\mbox{\boldmath$\sigma$}_i$ are Pauli spin operators for carriers
at site $i$,  and $J$ is the coupling constant between carriers
and core spins ${\bf S}_i$, the latter treated  classically as
unit length vectors ${\bf S}^2_i=1$. It should be understood from
the beginning that our aim is the study of transport properties of
the carriers, and not the ground state of  $H$. Therefore, we
consider  the random distribution of core spin {\em as given},
parameterizing the problem. Consistently with the discussion of
the introduction, we assume this probability distribution to be
site-uncorrelated, and chosen to describe the change from a
paramagnetic situation to a (weakly) ferromagnetic one. This can
be achieved, for instance, with a probability $ {\cal P} ({\bf S})
\propto \exp (h_{eff} S_z)$. Within this parametrization, $h_{eff}
= 0$ corresponds to the paramagnetic phase, and $h_{eff} \ne 0$
describes a phase where the core spins are polarized along the z
axis.

 Following standard practice, we define the magnetoresistance coefficient,
  ${\mathcal C}$, as:
\begin{equation}
\frac{\rho(0) - \rho(m)}{\rho(0)} = {\mathcal C} \: m^2 \label{C}
\end{equation}
where $\rho(m)$ is the resistivity corresponding to an average
polarization of core-spins $m = \langle S_z \rangle$. Notice that
eq. \ref{C} only makes sense to order $m^2$, the only situation we
will consider here.

 Before embarking on more complex formalisms, let us consider what perturbation
theory (in the coupling $J$) has to say about the magnetoresistance. In the standard
relaxation time approximation, the resistivity is given by:
\begin{equation}
\rho = \frac{m_b}{n e^2} \tau^{-1}
\label{rho}
\end{equation}
where $n$ is the number of carriers, $e$ the unit charge, $m_b$ the band mass, and $\tau$ is the
relaxation time, given to lowest order in the random potential by:
\begin{equation}
\tau^{-1} \propto  J^2(\langle {\bf S}^2 \rangle -  \langle {\bf
S} \rangle^2) \; {\cal D} (\epsilon_F) \label{tau}
\end{equation}
being ${\cal D} (\epsilon_F) $ the density of states per spin at the Fermi level.
 Upon magnetizing the core spins, only
changes in the fluctuating potential will modify the conductivity
to order $J^2$. Therefore, the relevant dependence is $\tau^{-1}
(m) \sim J^2 (1-m^2)$, leading to:
\begin{equation}
\frac{\rho(0) - \rho(m)}{\rho(0)} =
\frac{\tau^{-1}(0) - \tau^{-1}(m)}{\tau^{-1}(0)} = m^2.
\end{equation}
 We conclude that the magnetoresistance coefficient $\mathcal C$ is given by
\begin{equation}
{\mathcal C} = 1,
\end{equation}
a universal number, independent of the carrier concentration.
This uninteresting results seems to preclude further consideration of the weak
coupling limit, but, as we will show in the following sections, this is not the
whole story. We will obtain that, close to band edges, this perturbative result
fails, leading to an enhancement of $\mathcal C$ with decreasing carrier concentration.

\section{Coherent-Potential Approximation}
\label{cpa} The Hamiltonian of Eq. \ref{hamil} describes non
interacting electrons moving in a random potential. The obtention
of electronic properties requires, therefore, performing
configuration averages over the random orientation of core spins.
The Coherent-Potential Approximation (CPA)\cite{sov67,tay67}, a
procedure that we sketch here, offers a convenient way in order to
obtain such averages (see, for instance, ref.\cite{econom} for
coverage of the original  literature). We start with the first
term of Eq. \ref{hamil} as our unperturbed Hamiltonian
$H_o=\sum_{\bf{k},\sigma} \epsilon_{\bf{k}} \:
c_{\bf{k},\sigma}^{\dagger} c_{{\bf k},\sigma} $, leading to a
Green's function or resolvent whose local matrix elements are
given by:
\begin{equation}
g_o(z) = \langle i,\sigma|(z-H_o)^{-1}|i,\sigma \rangle =\int
d\epsilon \: \frac{{\cal D}_o(\epsilon)}{z-\epsilon}
\end{equation}
where $|i,\sigma \rangle$ is the electron state at site $i$ and
spin $\sigma$, and $z$ is a complex energy in the upper plane.
This unperturbed system is characterized by a reference density of
states (per spin) ${\cal D}_o(\epsilon)= -\pi^{-1} {\rm Im} \;
g_o(\epsilon + i0^+)$, which we take to be the usual
semielliptical function:
\begin{equation}
{\cal D}_o(\epsilon) = \frac{2}{\pi W} \sqrt{1-\left(\frac{\epsilon}{W}\right)^2}
\end{equation}
notice that band edges are entirely consistent with a 3-d system.

 The configuration average of the Green's function for the entire $H$:
 \hbox{$g_{\sigma}(z) = \langle \langle i,\sigma|(z-H)^{-1}|i,\sigma \rangle \rangle$} (external brackets stand for
configuration average) is calculated by the CPA
from that of the unperturbed system by means of a self-energy $\Sigma_{\sigma} $, in the following
manner:
\begin{equation}
g_{\sigma}(z) =g_o(z-\Sigma_{\sigma})
\end{equation}
where the self-energy, $\Sigma_{\sigma}(z) $, itself a function of
$z$, can be thought of as providing an effective medium that
substitutes the real system. This self-energy is obtained by
imposing that the local scattering produced when we replace an
effective site by a real one be, on the average, zero
\cite{econom}. Adapted to our case, the self-energy obeys the
following equation (rotational symmetry around the z axis
assumed):
\begin{equation}
\frac{1}{g_o(z-\Sigma_{\sigma})} = \left\langle
F_{\sigma}(z,\theta) - \frac{J^2
\sin^2\theta}{F_{-\sigma}(z,\theta)} \right\rangle
\end{equation}
where $\cos\theta = S_z$, brackets stand for averages over the distribution of
core spin orientations, and
the auxiliary function $ F_{\sigma}(z,\theta)$ is given by
\hbox{$F_{\sigma}(z,\theta) = \Sigma_{\sigma} + \sigma J \cos\theta +1 /
g_o(z-\Sigma_{\sigma})$},
where $\sigma=+1 (-1)$ for up (down) electron spin.

Although primarily intended for one-body properties (such as the
density of states), the CPA can be applied to transport
properties. The basic idea consists in decoupling the two-body
correlations that appear in the Kubo formula into products of
one-body correlators, which are then obtained with the CPA
self-energy. Skipping further details, the relevant expression
(see ref.\cite{econom}) adapted to our case is
\begin{equation}
\rho^{-1} = \sum_{\sigma}\rho^{-1}_{\sigma}
\end{equation}
where the conductivity for spin $(\sigma)$ carriers, $\rho^{-1}_{\sigma} $,
 is given by:
\begin{equation}
\rho^{-1}_{\sigma} \propto \int d\epsilon' f(\epsilon')
\left( Im \frac{1}{\epsilon_F+i0^+-\Sigma_\sigma(\epsilon_F+i0^+) - \epsilon'}
\right)^2
\end{equation}
where $\epsilon_F$ is the Fermi level and the function
$ f(\epsilon')\propto (1-(\epsilon'/W)^2)^{3/2}$ accounts for matrix
elements of the velocity operator\cite{econom,chatto00}.

 The whole CPA procedure for transport properties can be considered as a
 mean-field approximation. Actually a {\em double} mean-field: first for the
 resolvent and later for transport (decoupling). Therefore, its main shortcoming
 will be fluctuation related effects (i.e. localization). Nevertheless, we expect
 the 3-d nature of our problem to mitigate this limitation. In fact, the CPA has
 been shown to provide a good description of 3-d disordered systems, being one of
 the very few non perturbative methods available. In recent times, this
 approximation is often termed {\em dynamical mean field}\cite{Getal96}, an approach invented
 for genuine many body problems that becomes equivalent to the CPA when applied
 to a Hamiltonian like that of eq. (\ref{hamil}).

\section{Results. Weak coupling limit: $J/W << 1$}
\label{weak}
\begin{figure}
  \includegraphics[clip,width=8cm]{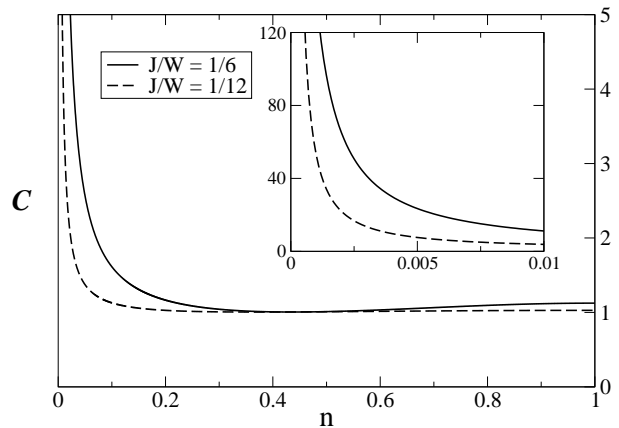} \\
  \caption{Magnetoresistance coefficient ${\mathcal C} $ (eq. \ref{C}) versus carrier
  density for two values of $J$. Inset: enlarged view of the low density
  region.}
\label{fig.C_weak}
\end{figure}

We have calculated the
magnetoresistance coefficient by direct numerical subtraction of the CPA
conductivities in the paramagnetic and ferromagnetic cases, the latter induced
with a tiny effective field in the probability distribution for core spins: $
{\cal P} ({\bf S}) \propto \exp (h_{eff} S_z)$. Representative results for
coupling constants well below the bandwidth  $(J <<W)$ are shown in
fig. \ref{fig.C_weak}, where the magnetoresistance coefficient $({\mathcal C})$ is
plotted versus carrier density per site $(n)$. Notice that, in this weak
coupling regime, the perturbative value ${\mathcal C}=1$ is closely approached
over most of the concentration range. Yet, significant enhancement of
${\cal C}$ is evident for small carrier concentration, that is, close to the band
edge. The inset of fig. \ref{fig.C_weak} blows up the low density regime,  where
values of ${\mathcal C} > 100$ are easily obtained. In fact, our numerical results
are compatible with a divergent behavior of ${\mathcal C}$ in that limit.
Fig. \ref{fig.C_weak}  is representative of the small coupling regime $J << W$:
no matter how weak $J$ is chosen, we can always find large enhancements of
${\mathcal C}$ if we approach the band edge. In fact, the smaller the value of
$J$, the smaller the range of densities required to see this enhancement, as
shown in the inset on fig. \ref{fig.C_weak}.  Notice that this effect does not
formally  contradict the perturbative result: for {\em fixed} carrier density
(i.e. fixed  Fermi level), ${\mathcal C}\rightarrow 1$ when  $J\rightarrow 0$.
Nevertheless, the quantitative failure of the perturbative results for a given
$J$ seems unavoidable upon approaching the band edges. In what follows, we will
address the origin of this failure with a simplified analytical procedure.

\subsection{Effect of band splitting. Analytical treatment}
\begin{figure}
  \includegraphics[clip,width=8cm]{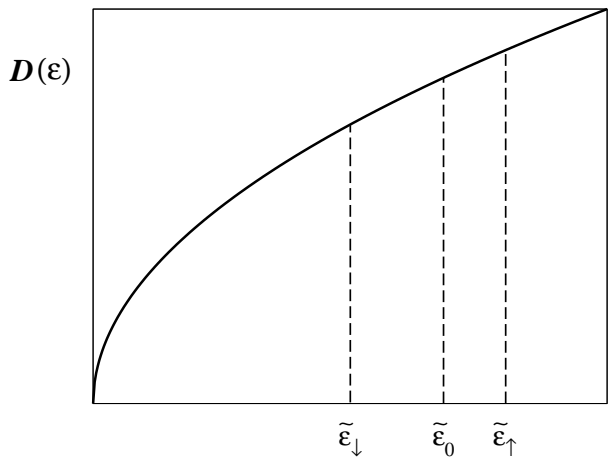} \\
  \caption{Schematic density of states showing the relative positions of Fermi
  levels upon magnetization.}
\label{fig.dens}
\end{figure}

 Our numerical  results
(and the analysis to follow) indicate that the nominal criterion
for the validity of the perturbative result, $J <<W $, valid well
within the band, has to be replaced by $J <<\tilde{\epsilon}$,
where $\tilde{\epsilon}= \epsilon_F-(-W) $, is the distance of the
Fermi level to the band edge. To see how this comes about and its
effect on the magnetoresistance, the original CPA equations are
rather opaque and inconvenient. Instead, we will start with the
perturbative treatment of section \ref{perturb}, adding the effect
we believe is at the origin of the enhancement: band splitting.
The first manifestation of a net polarization of core-spins, $m$,
is a band splitting of the unperturbed density of states ${\cal
D}_o(\epsilon \pm J m)  $. This splitting does not show up in the
resistivity if one keeps the calculation to  order $J^2$, but will
affect higher order terms. The situation is depicted in fig.
\ref{fig.dens}, where the up and down densities have been aligned
to share a common origin, leading to different {\em apparent}
Fermi levels for up and down electrons:
$\tilde{\epsilon}_{\uparrow} -  \tilde{\epsilon}_{\downarrow} = 2
m J$. Transport properties depend on the density of states at the
apparent Fermi levels which, in turn, changes rapidly upon
splitting (magnetization) close to a band edge. Therefore, it is
not unexpected that this mere shift can cause a large effect on
the magnetoresistance. We will see that this is the case.

We assume the simplified situation described in fig. \ref{fig.dens} with a parabolic band ${\cal
D}_o(\tilde{\epsilon}) \sim \sqrt{\tilde{\epsilon}} $, and apply standard transport
theory over this already split situation. Therefore, we have to discriminate between
up and down contributions to the conductivity:
\begin{equation} \label{rhosigma}
\rho^{-1} =  \sum_{\sigma}
\frac{n_{\sigma} e^2}{m_b} \tau_{\sigma}
\end{equation}
 where $ n_{\sigma}$, is
the number of carriers for each spin species, of course  satisfying the constraint
$ n=\sum_{\sigma}n_{\sigma}$. The inverse relaxation times are given by
\begin{eqnarray}
\nonumber
\tau_{\uparrow}^{-1} \propto J^2 \left[{\cal D}_o(\tilde{\epsilon}_{\uparrow})
  (\langle S_z^2 \rangle - m^2) +
{\cal D}_o(\tilde{\epsilon}_{\downarrow})
  \langle S_x^2+S_y^2 \rangle \right] \\ \nonumber
\tau_{\downarrow}^{-1} \propto J^2 \left[{\cal D}_o(\tilde{\epsilon}_{\downarrow})
  (\langle S_z^2 \rangle - m^2) +
{\cal D}_o(\tilde{\epsilon}_{\uparrow})
  \langle S_x^2+S_y^2 \rangle \right]
\end{eqnarray}
 where the
first contribution comes from transitions within the same spin species, and the
second term accounts for spin-flip transitions (notice the different density of
states in each case).

Going from the paramagnetic situation (Fermi level at
$\tilde{\epsilon}_o$  in fig. \ref{fig.dens}) to the ferromagnetic
one (Fermi levels at $\tilde{\epsilon}_{\uparrow,\downarrow}$),
the magnetoresistance picks up contributions coming from changes
in carrier spin populations $ n_{\sigma}$ and lifetimes
$\tau_{\sigma}$, in addition to the standard contribution already
described in section \ref{perturb}. Keeping consistently terms to
order $m^2$ in eq. (\ref{rhosigma}), it is only very tedious to
show that the magnetoresistance coefficient can be written as :
\begin{equation}\label{deltac}
{\mathcal C} = 1 + \frac{19}{36} \left( \frac{J}{\tilde{\epsilon}_o}\right) ^2
\end{equation}
We see that it is just a matter of getting close
enough to the band edge $\tilde{\epsilon}_o \leq J $ to obtain arbitrarily large
corrections to the nominal
perturbative result.

\begin{figure}
  \includegraphics[clip,width=8cm]{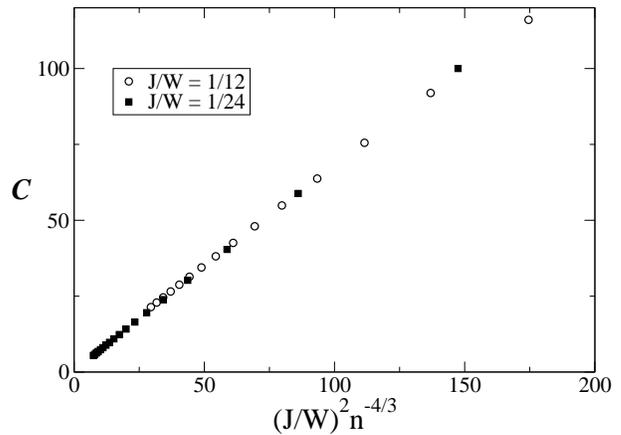} \\
  \caption{Magnetoresistance coefficient ${\mathcal C} $  versus carrier
  density for two values of $J$ illustrating the scaling behavior of
  eq. \ref{scaling} in the low density limit.}
\label{fig.scaling}
\end{figure}

In this simplified scheme, owing to the fact that
$ n\sim  \tilde{\epsilon}_o ^{\;3/2}$,
the magnetoresistance becomes divergent in the low carrier
density with the following law:
\begin{equation} \label{scaling}
{\mathcal C} -1 \sim \left( \frac{J}{W} \right)^2 n^{-4/3}
\end{equation}
 This scaling seems to be obeyed by the CPA results, as illustrated in
 fig. \ref{fig.scaling},
 where the result for two values of $J << W$ collapse onto the same straight line.
 This
 makes us believe that this simple treatment: splitting plus standard relaxation
  time
 transport, captures the essence of the enhancement observed in the CPA
 calculations.

\subsection{Ioffe-Regel criterion} We have seen that, for $J \ll W$, CPA
calculations and the simplified treatment of Eq. (\ref{deltac})
support the existence of a large magnetoresistance coefficient
close to the band edge. In this section we address the validity of
this result. The main concern comes from the fact that we have to
be close to a band edge, where localization effects (ignored in
our mean-field approach) are expected to play an important
role\cite{and58,mott,econom}. To estimate the energy range of this
effect, we use the Ioffe-Regel criterion\cite{iof60,mott}. This
criterion can be stated saying that quantum corrections to
transport can be expected to be relevant when the mean free path
diminishes to become of the order of the de Broglie wavelength for
electrons at the Fermi level. This sets a characteristic  distance
to the band edge $\tilde{\epsilon}_{IR}$ such that, if the Fermi
level is below it, $ \tilde{\epsilon}_o < \tilde{\epsilon}_{IR} $,
localization is expected to dominate. In our case, this criterion
can be written as the condition:
\begin{equation}
\tau^{-1}_{IR} \sim \tilde{\epsilon}_{IR},
\end{equation}
where $\tau_{IR}$ is the lifetime at $\tilde{\epsilon}_{IR}$.
Using the perturbative result for the lifetime, this leads to:
\begin{equation} \label{ioffe}
\tilde{\epsilon}_{IR} \sim 8 W \left( \frac{J}{W} \right)^4
\end{equation}
Comparing this $J^4$ dependence with the $J^2$  of Eq.
(\ref{deltac}), we see that there is ample room for observing the
enhancement of ${\mathcal C}$ upon approaching the band edge,
before the Ioffe-Regel limit is reached. More quantitatively, we
can define the  enhancement factor at the Ioffe-Regel limit by:
\begin{equation} \label{ioffe1}
\Delta{\mathcal C}_{IR} = \frac{19}{36} \left( \frac{J}{\tilde{\epsilon}_{IR} }\right) ^2
\end{equation}
with the result that:
\begin{equation}
\Delta {\mathcal C}_{IR} \sim 10^{-2} \left(\frac{W}{J} \right)^6
\end{equation}

Therefore, we conclude that, for our weak coupling case $J \ll W$,
large increases of the magnetoresistance coefficient $\mathcal C $
are allowed before localization effects set in.

\section{Results. Double exchange limit: $J/W \gg 1$}
\label{strong}
\begin{figure}
  \includegraphics[clip,width=8cm]{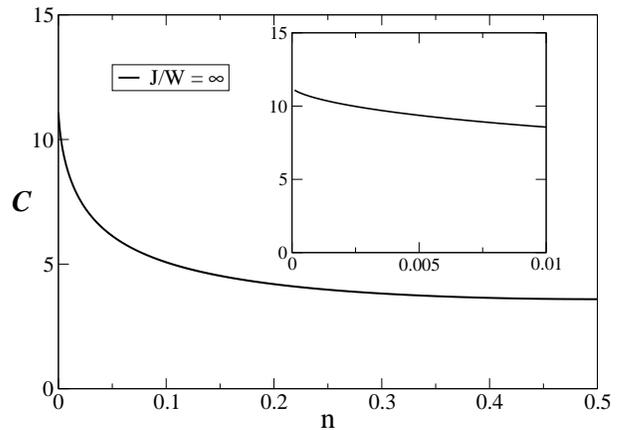} \\
  \caption{Magnetoresistance coefficient ${\mathcal C} $ (eq. \ref{C}) versus carrier
  density for $J/W = \infty$. Inset: enlarged view of the low density
  region.}
\label{fig.C_strong}
\end{figure}

 In this limit, the
splitting between the two spin subbands is large, and we can neglect the one
which is at high energy, $\sim J$. We have implemented the $J=  \infty $ limit
in the CPA equations. A similar calculation has been carried out before by
Furukawa\cite{F94}, but here we are mainly interested in results close to the
band edge. In fig. \ref{fig.C_strong} we show the CPA results for the
magnetoresistance coefficient ${\mathcal C} $ as a function of carrier density.
Notice a mild enhancement of ${\mathcal C}$ close to the band edge but, unlike the
weak coupling limit, no divergent behavior is observed. This is more clearly
seen in the inset of fig. \ref{fig.C_strong}, where the low density region is
blown up, showing   a saturation of the magnetoresistance coefficient around
${\mathcal C}\sim 11 $.

 Direct use of the CPA equations to  understand this behavior is, again,
inconvenient. A simpler picture in order to include the effect of a finite
magnetization is to assume that the bandwidth depends on $m$\cite{G60} in the
following manner: $W_m = W \sqrt{(1+m^2)/2} $.  This ansatz interpolates
between the paramagnetic $(W_m = W / \sqrt{2})$ and fully ferromagnetic limit
$(W_m = W)$, giving a reasonable approximation to the dependence of the Curie
temperature on band filling\cite{AGG99,GGA00}. Assuming a semielliptical
density of states and measuring energies from the center of the occupied band,
we have:
\begin{equation}
{\cal D}(\epsilon) = \frac{2}{\pi W_m}
\sqrt{1-\left(\frac{\epsilon}{W_m}\right)^2},
\end{equation}
and the change in the electronic Green's function can be described by
a self-energy, $\Sigma ( z, m )$, such that:
\begin{equation}
g(z,m) =  g_o[z - \Sigma (z , m ) ]
\end{equation}
with $z=\epsilon + i0^+ $. This self-energy is:
\begin{equation}
\Sigma ( z , m ) = \frac{1 - m^2}{2 ( 1 + m^2 )} \left[ - z
 + \sqrt{z^2 - W_m^2} \right]
\end{equation}
We now assume that the conductivity  is proportional to $n
\times {\rm Im} \Sigma ( \epsilon_F , m )^{-1}$ as done in
previous sections. Then:
\begin{equation}
{\mathcal C} \sim - \frac{1}{{\rm Im} \Sigma ( \epsilon_F , 0 )}
\left. \frac{\partial^2 {\rm Im} \Sigma ( \epsilon_F , m )}
{\partial m^2} \right|_{m^2 \rightarrow 0}
\end{equation}
 In this equation
we have to take into account that, if the number of particles is
fixed,  $\epsilon_F \approx W \sqrt{(1 + m^2)/2} ( -1 + [ ( 3 \pi ) /
( 4 \sqrt{2} ) ]^{2/3} n^{2/3} )$ has an implicit dependence on
$m^2$. Then ${\rm Im} \Sigma \sim n^{2/3} W ( 1 - m^2 ) / \sqrt{1 + m^2}$,
and we find no divergent term in ${\mathcal C}$ as $n \rightarrow
0$. Therefore,
\begin{equation}
\lim_{n \rightarrow 0} {\mathcal C} \sim {\rm constant}
\end{equation}
This saturation of ${\mathcal C}$ is obtained in the full CPA
calculation (see fig. \ref{fig.C_strong}), lending support to this
simple picture of a $m$-dependent bandwidth in the limit
$J=\infty$. Notice that, for this picture to apply, the number of
carriers must be kept fixed upon magnetization. If  the chemical
potential were  kept fixed, on the other hand, we would have
obtained a divergent behavior: $\lim_{n \rightarrow 0} {\mathcal
C} \sim n^{-2/3} $.
\begin{figure}
  \includegraphics[clip,width=10cm]{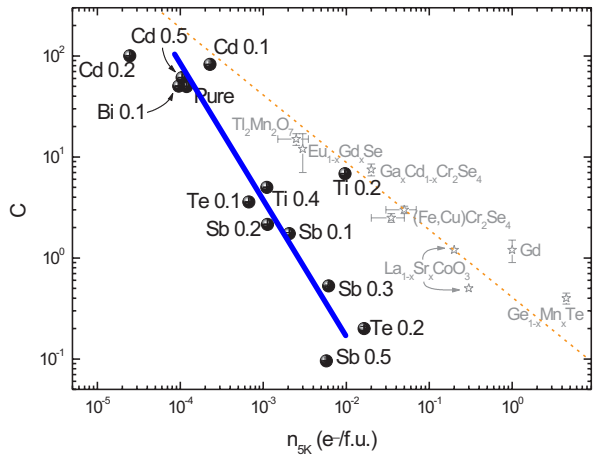} \\
  \caption{Value of ${\mathcal C}$ for the pyrochlore compounds analyzed
  in\protect{\cite{Vetal02,V02}} (dark dots) and for the magnetic materials
  studied in\protect{\cite{ML98}} (stars). The full line is
  a fit to a $n^{-4/3}$ dependence. The broken line is the $n^{-1/3}$ dependence
  discussed
  in\protect{\cite{ML98}}.}
\label{C_exp}
\end{figure}

\section{Validity of the approximations. Summary.}
\label{summary}
The analysis discussed above uses an effective medium approach to
study the transport properties of itinerant electrons coupled to
classical spins. As mentioned in the introduction, the method
neglects the spatial correlations of the magnetic fluctuations: a potentially
serious drawback if the source of polarization is the spontaneous magnetic order
expected for the Hamiltonian of eq. \ref{hamil}.
The critical behavior of these fluctuations is particularly important
in the vicinity of the point ${\bf \vec{q}} =
0$. In addition, our scheme cannot take into account the
localization of the electronic states induced by the disorder in
the magnetization near the band edges. Both limitations become
important when the electronic density is lowered, as the relevant
scattering processes are shifted towards low momenta, and the
Fermi energy approaches the band edge. We will consider these two
effects separately.

The lack of spatial correlations is a generic feature of effective
medium theories. These correlations modify the momentum dependence
of the susceptibilities at low values of the momenta. If the
number of dimensions of the system were indeed large, these
effects could be safely ignored, as the volume of the region near
${\bf \vec{q}} = 0$ is negligible compared to the volume of the
unit cell. We are interested in changes in the resistivity near
the Curie temperature, so that critical fluctuations at low
momenta are always present. The length scale at which these
fluctuations become important is the correlation length in the
paramagnetic phase, $\xi_0 $, whereas the length scale for
transport is $k^{-1}_F$\cite{ML98b}. Therefore, we expect our
results to be applicable for $k_F \xi_0 \gg 1$, leading to the
following condition for the carrier density: $n \gg (a/\xi_0)^D$,
where $D$ is the dimension and $a$ the lattice size.  Assuming,
for instance, a typical value $\xi_0 = 5a$ and $D = 3$, we find $n
\geq 10^{-2}$.

Localization  becomes important when the mean free path associated to the
disorder, $l$ is such that $k_F l \sim 1$.  In the weak coupling limit, its
effect has been discussed before, with conclusions that we briefly repeat
here. Localization is expected to be relevant below the Ioffe-Regel energy, $
\tilde{\epsilon}_{IR}\sim J^4 / W^3$.  Therefore, observing the enhanced
behavior ${\cal C} \sim n^{-4/3}  $ requires the following condition on the
Fermi level:
\begin{equation}
\frac{J^4}{W^3} \ll \tilde{\epsilon}_F \ll J,
\end{equation}
a regime easily accessible in the weak coupling limit, $J \ll W  $.

In the double exchange limit, $J/W \gg 1$, we have ${\rm Im}
\Sigma \sim \sqrt{W \tilde{\epsilon}_F} \ge \tilde{\epsilon}_F$
for $n \ll 1$. The mean free path becomes comparable to the
lattice spacing at low densities, implying that localization
effects are relevant in this regime. Notice, though, that the
applicability of the previous criterion, imported from the
perturbative limit, is open to question in this strong coupling
limit\cite{varma96}. Moreover, the use of the scheme used here to
study the phase diagram shows that the paramagnetic-ferromagnetic
transition becomes first order for $n \leq 10^{-1}$\cite{AGG99}.
This inclusion of this effect can change significantly the
magnetoresistance.

In summary, we have studied transport properties of electrons
coupled to local, uncorrelated core spins. The magnetoresistance
coefficient (eq. \ref{C}) has been shown to diverge in the weak
coupling limit as ${\cal C}\sim n^{-4/3}$, near the band edge, as
a consequence of band splitting upon magnetization. On the other
hand, the magnetoresistance remains finite in the strong coupling
limit (double exchange model) close to the band edge. This is
understood from a simplified picture consisting in a
magnetization-modulated bandwidth. Possible limitations of the
previous results from the neglect of both localization effects and
critical fluctuations have been addressed.

Finally, we would like to consider the potential applicability of
our results to the experimental situation. In Fig.[\ref{C_exp}],
we present available information\cite{Vetal02,V02} for the
magnetoresistance in pyrochlores and other magnetic materials. We
observe a strong dependence on the density for the pyrochlores
family, which seems to come closer to the $n^{-4/3}$ law obtained
in this paper than the fit based on $2k_F$ scattering alone
\cite{ML98}
\section{Acknowledgements.}
We are thankful to J. L. Mart{\'\i}nez for many helpful
conversations, and to P. Velasco and J. L. Mart{\'\i}nez for
sharing with us the data shown in Fig.[\ref{C_exp}]. Financial
support from MCyT (Spain), through grant no. MAT2002-04095-C02-01
is gratefully acknowledged. Additional support was provided
through the France-Spain binational grant PAI PICASSO 05253WF.

\bibliography{guillermo}

\end{document}